\documentclass[aps,twocolumn,showpacs,amsmath,amssymb]{revtex4}
\usepackage{dcolumn}
\usepackage{bm}
\usepackage{color}
\usepackage{latexsym}
\usepackage{amssymb}
\usepackage{graphicx}
\usepackage{ulem}

\begin{document}
\title{On the Quantum Hall Effect in graphene}

\author{M. V. Cheremisin}

\affiliation{A.F.Ioffe Physical-Technical Institute, 194021
St.Petersburg, Russia}

\begin{abstract}
Quantum Hall effect in 1,2-layer graphene is analyzed. The transverse and longitudinal resistivity
are found to be universal functions of the filling factor and temperature. At fixed magnetic field mode the
magneto-transport problem is resolved in the vicinity of the Dirac point.
\end{abstract}


\maketitle

\section{\label{sec:Introduction}Introduction}
Recently, a great deal of interest has been focused on the
electric field effect and transport in a two-dimensional electron gas(2DEG) system formed in graphene flake \cite{Novoselov04}. In the present paper, we are are mostly concerned with the magnetotransport in monolayer and bilayer graphene in a strong magnetic field.

\section{\label{sec:Electric Field Effect in monolayer graphene at B=0} Electric Field Effect in graphene at B=0}
\begin{figure}
\begin{center}
\includegraphics[scale=0.75]{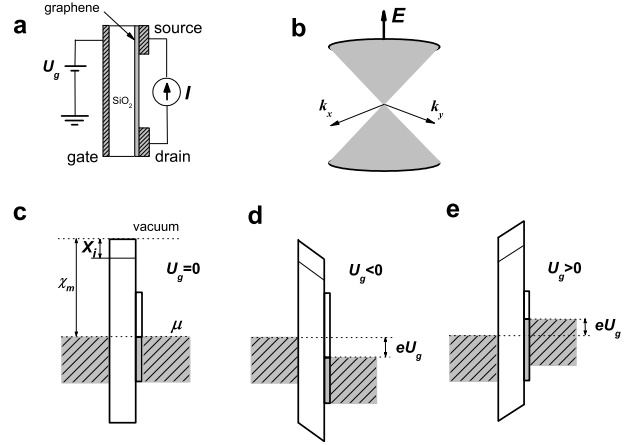} \caption[]{\label{Fig1}(a)Experimental setup for gated graphene.
(b)Band structure at $\bold{k} \simeq 0$, which shows the Dirac cones. The energy diagram for (c)$U_{g}=0$,
(d)$U_{g}<0$ and (e)$U_{g}>0$.  $\chi_{m}$ is the metal work function, and $X_{i}$ the electron affinity of the insulator.}
\end{center}
\end{figure}

We first focus on the transport properties of single-layer graphene (SLG) at B=0. The typical experimental setup is shown in Fig.\ref{Fig1}, a. The source and drain terminals of the graphene sample are connected to the current source. The metal backgate and the drain are connected via the source of voltage, $U_{g}$, serving to change the  carrier density in graphene. At low-current mode $I \rightarrow 0$, the voltage drop, $U_{sd}$, between the source and drain contacts is measured, and, hence, it can be used to determine the graphene resistance.

According to Refs.\cite{Wallace47},\cite{McClure56}, the SLG energy spectrum obeys the linear dependence
\begin{equation}
E_{\pm}(k)=\pm E(k)=\pm \hbar \upsilon k, \label{SLG spectrum_B0}
\end{equation}
where $E(k)$ is energy, and $k$, the distance in the $\bold{k}$-space relative to the zone edge ( see Fig.\ref{Fig1}, b ). The "$\pm$" sign refers to the electron (+) and hole (-) conducting  bands, respectively. The SLG state $E=0$ is named the Dirac point(DP). It will be recalled that the Fermi energy, $\mu$, in graphene can be varied either by means of the backgate voltage via field-effect\cite{Novoselov04} or by chemical doping. When $\mu > 0$ ($\mu < 0$), the Fermi level falls within the electron ( hole ) conducting band, respectively. For the special case, $\mu=0$, in which the Fermi level coincides with the Dirac point, the density of conducting electrons is equal to that of holes.

At low currents, the voltage drop across the graphene sample is small as compared to applied gate voltage. Neglecting the voltage difference between the source and drain terminals,
we plot in Figs.\ref{Fig1}, c-e the transverse energy diagram for an arbitrary gate bias. At a fixed gate voltage, $U_{g}$, the quasi-Fermi level of the backgate contact is shifted by the energy, $eU_{g}$, with respect to that of the drain contact and, respectively, to the quasi-Fermi level, $\mu$, of the graphene monolayer. Actually, the quasi-Fermi level of graphene sample is, pinned by that of the drain electrode.

At the same time, the combined structure in Fig.\ref{Fig1}, a can be regarded as a capacitor in which the graphene monolayer plays the role of one of its plate. At a fixed gate voltage $U_{g}$, the
charge density in the graphene monolayer is $Q=CU_{g}$, where $C=\epsilon_{0}\epsilon /d$ is the capacitance per unit area; $d$, gate thickness; $\epsilon_{0}$ and, $\epsilon$, permittivity of free space and the relative permittivity of the SLG substrate, respectively. It is noteworthy that, at a certain Fermi energy, graphene is in charge-neutrality ( CN ) state when $Q,U_{g}=0$. Without chemical doping, the state in graphene, associated with the Dirac point, coincides with the CN state. Doping shifts the DP state with respect to the CN state. The gate bias required to reach the DP state is called the offset voltage. For simplicity, we further neglect the chemical doping.

\begin{figure}
\begin{center}
\includegraphics[scale=0.75]{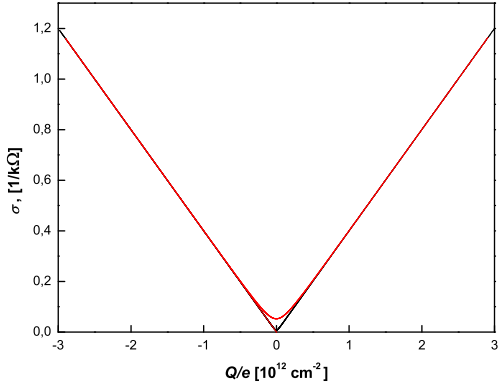} \caption[]{\label{Fig2} Conductivity of SLG at
T=100;273K and $\mu_{eh}=2.5 \times 10^{3}$cm$^{2}/Vs$ vs. the charge density.}
\end{center}
\end{figure}

Using the Gibbs statistics, we can distinguish the components of the thermodynamic potential for electrons,
$\Omega_{e}$, and holes, $\Omega_{h}$, and, then represent them as follows
\begin{eqnarray}
\Omega_{e} =-kT\sum \limits_{k} \ln \left(1+e^{\frac{\mu -E(k)
}{kT}}\right),
\label{SLG_OMEGA}\\
\Omega_{h}=\Omega_{e}(-\mu),
\nonumber
\end{eqnarray}
which gives the electron (hole) concentration as:
\begin{equation}
N=- {\partial \Omega_{e}
\overwithdelims()\partial \mu }_{T}, P={\partial \Omega_{h}
\overwithdelims()\partial \mu }_{T}.
\label{NP_DEFINITION}
\end{equation}

Using the SLG density of states, $D(E)=\frac{2|E|}{\pi \hbar ^{2}\upsilon^{2}}$, which includes both
the valley and spin degeneracies, we obtain
\begin{equation}
N=N_{0}2\xi^{2}F_{1}(1/\xi ), P=N_{0}2\xi^{2}F_{1}(-1/\xi ),
\label{SLG_N_B0}
\end{equation}
where $\xi =kT/\mu$ is the degeneracy parameter, and $F _{n}(z)$, the Fermi integral. Then,
$N_{0}=\frac{1}{\pi}\left( \frac{\mu}{\hbar \upsilon}\right )^{2}$ is the density of degenerate
electrons (holes) at $T=0$. It is noteworthy that, at fixed $\mu$ and finite temperature, the electron (hole) density exceeds $N_{0}$. For example, for Dirac point state we
obtain the intrinsic electron (hole) densities as
$N_{i}=\frac{\pi}{6}\left( \frac{kT}{\hbar \upsilon}\right )^{2}$. It is instructive to estimate the typical Fermi energies and carrier densities. For the Fermi energy $\mu=0.2$eV and  $\upsilon=8 \times 10^{7}$ñm/s, we obtain the electron density as $N_{0}=4.6 \times 10^{12}$cm$^{-2}$. The temperature range $T=4-273$K corresponds to a degeneracy parameter $\xi=0.002-0.11$. At $T=100$K, the intrinsic density
is $N_{i}=7 \times 10^{9}$cm$^{-2}$.

We now can calculate the SLG charge density, $Q$, and the conductivity, $\sigma$, as follows:
\begin{equation}
\sigma=e\mu_{eh}(N+P),\qquad \label{SLG_SIGMA_B0}
Q=e(N-P).
\end{equation}
For simplicity, we further assume a constant mobility, $\mu_{eh}$, for both kinds of carriers.
In Fig.\ref{Fig2}, we plot the dependence of the conductivity on the SLG charge density $Q/e= CU_{g}/e$. As expected, the curve $\sigma(Q)$ is symmetric and, moreover, it exhibits the minimum conductivity,
$\sigma_{min}=2e\mu_{eh} N_{i} \sim T^{2}$, at the Dirac point. In the vicinity of the DP, when $| \xi | \ll 1$, the conductivity $\sigma=\sigma_{min}\left(1+a\frac{Q^{2}}{eN_{i}^{2}}\right )$, where $a \sim 0.1$ is a coefficient. For the strongly degenerate case $|\xi | \gg 1$, we obtain $\sigma=\mu_{eh} | Q | \sim | U_{g} |$.

\section{\label{sec:QHE}IQHE in graphene}
\subsection{\label{sec: General formalism}General formalism }
We now follow the argumentation first put forward in Refs.\cite{Kirby73,Cheremisin01}, and then
modified\cite{Cheremisin05a} regarding the magnetotransport problem of 2D electron systems in strong magnetic fields, including Shubnikov-de Haas Oscillations(SdHO) and Integer Quantum Hall
Effect(IQHE) modes. In particular, this approach allows us to reproduce the SdHO beating pattern\cite{Cheremisin05b} known to occur at a finite zero-field valley splitting (Si-MOSFET 2D system) and in both the crossed- and tilt-field configuration case.

In the presence of a magnetic field, the current density, $\mathbf{j}$, and the energy flux density,
$\mathbf{q}$, are given by
\begin{equation}
\mathbf{j}=\hat{\sigma }(\mathbf{E-}\alpha \nabla
T),\quad \mathbf{q}=\left( \alpha T-\zeta /e\right)
\mathbf{j}-\hat{\kappa }\nabla T. \label{TRANSPORT}
\end{equation}
Here, $\mathbf{E}=\nabla \zeta /e$ is the electric field; $\zeta
=\mu -e\varphi $, electrochemical potential; $\nabla T$, temperature gradient; and $\alpha$,
thermoelectric power. Then, $\hat{\sigma }$ and $\hat{\kappa }=LT\hat{\sigma }$ are the conductivity
and thermal conductivity tensors respectively, and $L=\frac{\pi^{2}k^{2}}{3e^{2}}$ is the Lorentz number.
Equation (\ref{TRANSPORT}) is valid for a confined-topology sample, for which the diamagnetic surface
currents \cite{Obraztsov64} are taken into account. Both the Einstein and Onsager relationships are satisfied.

\begin{figure}
\begin{center}
\includegraphics[scale=0.75]{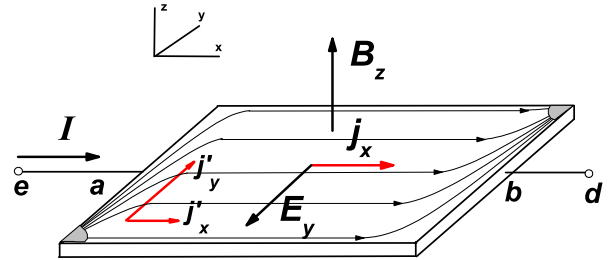} \caption[]{\label{Fig3} Experimental setup. The red arrows show the current flowing at the high-field corner and in the sample bulk.}
\end{center}
\end{figure}

As first demonstrated in Ref.\cite{Kirby73}, the longitudinal current $I_{x}=I$
flowing in the circuit with an embedded Hall-bar sample(see Fig.\ref{Fig3}) causes a certain temperature
gradient $\nabla_{x} T$ in the sample bulk. The gradient is due to the heating(cooling) at the first(second) sample contact, caused by the Peltier effect. Using Eq.(\ref{TRANSPORT}), for a homogeneous sample, $\nabla \mu=0 $, we obtain the $x,y$-components of the electric field:
\begin{eqnarray}
E_{x}-\alpha \nabla_{x} T =\rho_{xx} j_{x},
\label{TRANSPORT_B} \\
E_{y}=\rho_{yx}j_{x}.
\nonumber
\end{eqnarray}
Here, we assume that the transverse current and temperature gradient are absent in the sample bulk, i.e.
 $j_{y},\nabla_{y} T=0$. Then, $\rho_{xy(x)}=\frac{\sigma_{xy(x)}}{\sigma_{xx}^{2}+\sigma_{yx}^{2}}$
are the components of the resistivity tensor $\hat{\rho }=\hat{\sigma}^{-1}$.

We now calculate the longitudinal temperature gradient $\nabla_{x} T$. The
Peltier heat is known to be generated by a current
crossing the contact between two different conductors. At the
contact (e.g., contact "a" in Fig.\ref{Fig3}), the
temperature $T_{a}$, electrochemical potential $\zeta $, normal
components of the total current $I$, and the total energy flux
are continuous. By contrast, there exists a difference $ \Delta \alpha =\alpha
_{m}-\alpha$ between the thermoelectric powers of the metal and the sample,
respectively. For $\Delta \alpha >0$, the charge crossing
contact "a" gains an energy $e\Delta \alpha T_{a}$. Consequently,
$Q_{a}=I\Delta \alpha T_{a}$ is the amount of the Peltier heat released
per unit time in the contact "a". For $\Delta \alpha >0$ and
the current flow direction shown in Fig.\ref{Fig3}, contact "a" is
heated and contact "b" is cooled. These contacts are at
different temperatures, and $\Delta T=T_{a}-T_{b}>0$. At small
currents, the temperature gradient is small and $T_{a,b}\approx
T$. Therefore, the thermoelectric power of the sample, $\alpha $, can be assumed to be constant,
and we can disregard the Thomson heating in the sample bulk, $IT\nabla \alpha=0$.
It is noteworthy that, for conventional 2DEG, the cooling caused by the heat leak
via the leads and the 3D substrate is negligible\cite{Cheremisin05a} in quantizing magnetic fields.
In fact, we can consider the 2D sample under adiabatic cooling conditions, when the amount of the
Peltier heat released at contact "a" is equal to that absorbed
at contact "b". According to Eq.(\ref{TRANSPORT}) the energy flux $\mathbf{q}$ is continuous at each contact, and, therefore, we obtain the $x,y$-components of the energy flux in the form:
\begin{eqnarray}
\kappa_{yx}\left. \nabla _{x}T\right| _{a,b}=-j'_{y}\Delta \alpha T_{a,b},
\label{ENERGY_TRANSPORT_CONDITION} \\
\kappa_{xx}\left. \nabla _{x}T\right| _{a,b}=-j'_{x}\Delta \alpha T_{a,b}.
\nonumber
\end{eqnarray}
Here, we take into account that, in a finite magnetic field, the current is known\cite{Thompson70} to enter and leave the sample at two diagonally opposite high-field corners( spots ) of the Hall-bar sample( Fig.\ref{Fig3}). Accordingly, $j'_{x(y)}$ are the $x-y$ components of the current density in the vicinity of the high-field corner. Evidently, the total current that enters(leaves) the sample at
high-field spots, $I'=\sqrt{(I'_{x})^{2}+(I'_{y})^{2}}$, is equal to the current in the sample bulk, i.e. $I'=I=j_{x}d$, where $d$ is the sample width.
Folding Eq.(\ref{ENERGY_TRANSPORT_CONDITION}), we finally obtain the longitudinal temperature gradient
$\nabla_{x}T=-\frac{j_{x}\Delta \alpha}{L\sqrt{\sigma_{yx}^{2}+\sigma_{xx}^{2}}}$, which is linear in current.

Using Eq.(\ref{TRANSPORT_B}), the voltage drop $U$, measured between the open ends ''e'' and ''d''
can be easily found as $U=\int E_{x}dx=\Delta \alpha(T_{a}-T_{b})+\rho_{xx}j_{x}l$, where $l$ is the sample
length. With the resistance and the thermoelectric power of the metal leads disregarded, the \textbf{total}
resistivity $\rho=U/j_{x}l$ and the Hall resistivity $\rho_{H}=E_{y}/j_{x}$ of the sample are given by
\begin{equation}
\rho = \rho_{xx}+\frac{\alpha^2}{\sigma_{yx} L}\sin \theta_{H}
, \qquad \rho_{H}=\rho_{yx},
\label{RESISTIVITY}
\end{equation}
where $\theta_{H}=\arctan(\sigma_{yx}/\sigma_{xx})$ is the conventional Hall angle. It is noteworthy that the total resistivity
is an even function of the magnetic field. Equation (\ref{RESISTIVITY}) is the central result of the present paper. First, at zero magnetic field, we obtain $\sigma_{yx},\theta_{H}=0$, and, therefore, Eq.(\ref{RESISTIVITY}) reproduces the result $\rho = \sigma^{-1}(1+\alpha^2/L)$ reported by Kirby and Laubitz \cite{Kirby73}.

The primary concern of the present paper is the case of strong magnetic field, in which the cyclotron energy exceeds both the thermal energy and the energy related to the LL-width.  2D carriers can be considered
dissipationless\cite{Baskin78} in this case, and, therefore, we assume $\sigma_{xx}, \rho_{xx} \simeq 0$
and thus $\theta_{H}=\pi/2$. Equation (\ref{RESISTIVITY}) yields the result
\begin{equation}
\rho=\frac{\alpha^2}{\mid \sigma_{yx}\mid L}, \qquad \rho_{H}=\sigma_{yx}^{-1}.
\label{RESISTIVITY_HIGH_B}
\end{equation}
previously reported\cite{Cheremisin05a} for the unipolar magneto-transport by 2D electrons or holes.

In contrast to the conventional 2D systems, carriers of both kinds, electrons and holes, coexist in graphene \cite{Novoselov04,Wiedmann11}. Therefore, Eq.(\ref{RESISTIVITY_HIGH_B}) is only valid when the current-carrying state is provided by the majority carriers(electrons $ N \gg P$ or holes $N \ll P$). We emphasize that in strong magnetic fields the Hall conductivity of the electron-hole plasma
\begin{equation}
\sigma_{yx}=\frac{(N-P)ec}{B},
\label{SIGMAxx}
\end{equation}
may be low enough and, moreover, it vanishes at the Dirac point when $N=P$. The Hall angle vanishes as well, $\theta_{H} \rightarrow 0$. Actually, the current flow at DP is uniform throughout the sample, including the contacts. The high-field corners disappear. In the ultra-narrow vicinity of the DP, i.e., at $\sigma_{yx} \ll \sigma_{xx}$ Eq.(\ref{RESISTIVITY}) yields
\begin{eqnarray}
\rho = \sigma_{xx}^{-1}(1+\alpha^2/L), \qquad \rho_{H}=\frac{\sigma_{yx}}{\sigma_{xx}^{2}}.
\label{RESISTIVITY_DP}
\end{eqnarray}
According to Eq.(\ref{RESISTIVITY_DP}) the total resistivity $\rho$
is finite, whereas the Hall resistivity vanishes in agreement with experiments\cite{Novoselov04}.

\subsection{\label{sec: IQHE in SLG at variable magnetic field}IQHE in SLG at variable magnetic field }
In a strong magnetic field, the SLG spectrum is given by\cite{McClure56}:
\begin{equation}
E_{\mathcal{N}}^{\pm}=\pm E_{\mathcal{N}}= \pm \frac{\hbar \upsilon}{\lambda_{B}}\sqrt{2\mathcal{N}},
\label{SLG_SPECTRUM_B}
\end{equation}
where $\lambda_{B}=\sqrt{\hbar c/eB}$ is the magnetic length, and $\mathcal{N}=0,1,2..$, the Landau level(LL)
number. Here, the $\pm$ sign refers to the electron(+) and hole(-) energy bands, respectively.
We assume further no LL broadening.

With the energy spectrum specified by Eq.(\ref{SLG_SPECTRUM_B}), the electron(e) and hole(h)
part of the thermodynamic potential $\Omega_{e,h}$ are given by
\begin{eqnarray}
\Omega_{e}=-kT \Gamma \left[4\sum \limits_{N=1}^{\infty}\ln
(1+e^{\frac{\mu -E_{\mathcal{N}}}{kT}})+ 2\ln(1+e^{\frac{\mu}{kT}})\right],
\label{SLG_OMEGA_B}\\
\Omega_{h}=\Omega_{e}(-\mu, T),
\nonumber
\end{eqnarray}
where $\Gamma=(2\pi \lambda_{B}^{2})^{-1}$ is the zero-width LL density of states. In Eq.(\ref{SLG_OMEGA_B}),
we take into account the four-fold degeneracy( spin+valley ) for high-index ($\mathcal{N} \geq 1$) LLs. Then, the
zeroth LL exhibits\cite{McClure56} a double degeneracy for both electrons and holes.

Let us introduce the dimensionless electron energy spectrum $\varepsilon_{\mathcal{N}}=E_{\mathcal{N}}/\mu =\text{sign}(\nu)
\sqrt{4\mathcal{N}/|\nu|}$, where $\nu= \text{sign}(\mu)N_{0}/\Gamma$ is the conventional filling factor. We now consider as a starting point the magneto-transport solely by electrons ($N \gg P$)
at a finite temperature and certain Fermi energy $\mu=\hbar \upsilon \sqrt{\pi N_{0}} > 0$.  Using the thermodynamic definition specified by Eq.(\ref{NP_DEFINITION}) and Eq.(\ref{SLG_OMEGA_B}), we finally obtain the electron density
\begin{equation}
N=\Gamma \left[ 4\sum \limits_{\mathcal{N}=1}^{\infty}\mathcal{F}\left(\frac{\varepsilon_{\mathcal{N}}-1}{\xi}\right)+
2\mathcal{F}\left(\frac{-1}{\xi}\right)\right],
\label{SLG_N}
\end{equation}
where $\mathcal{F}$ is the Fermi function.

It will be recalled that, in a strong magnetic field, the thermoelectric power is known\cite{Obraztsov64,Girvin82} to be proportional to the entropy $S_{e}$ per particle. For electrons, we obtain
\begin{eqnarray}
\alpha_{e}=-{\frac{S_{e}}{eN}}, \label{ALPHA_DEFINITION_ELECTRON} \\
S_{e}=-{\partial \Omega_{e} \overwithdelims()\partial T }_{\mu}=\frac{4\Gamma}{T} \sum \limits_{E_\mathcal{N}}
\int_{E}^{\infty}(\mu-E)\mathcal{F'}dE.
\nonumber
\end{eqnarray}
It is noteworthy that the thermodynamic potential $\Omega_{e}$, density $N$, and thermoelectric power $\alpha_{e}$
are universal functions of the dimensionless magnetic field $\nu^{-1}$ and temperature $\xi$. In the case of low
temperatures and weak magnetic fields $\nu ^{-1},\xi \ll 1$, we use Lifshitz-Kosevich formalism and, thereby,
derive in Appendix \ref{Lifshitz-Kosevich formalism} asymptotic formulas for $\Omega_{e},N,S_{e}$, and, hence $\alpha_{e}$.

\begin{figure}
\begin{center}
\includegraphics[scale=0.75]{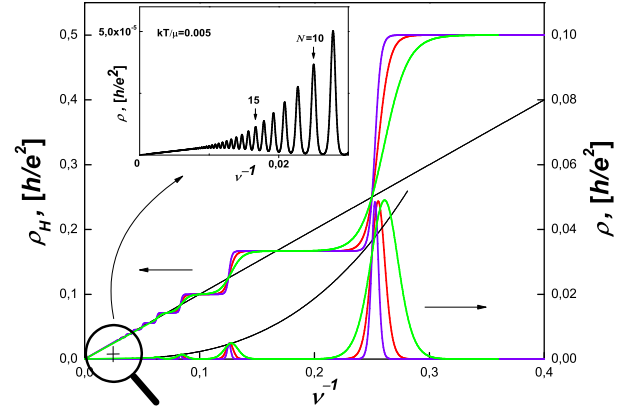} \caption[]{\label{Fig4} Magnetic field dependence of the total and Hall resistivities, calculated using the energy spectrum Eq.(\ref{SLG_SPECTRUM_B})
and Eqs.(\ref{RESISTIVITY_HIGH_B},\ref{SLG_N},\ref{ALPHA_DEFINITION_ELECTRON}) for electrons($\mu>0$) at $\xi=0.005, 0.01, 0.02$. The thin line represents the low-B asymptote $\rho_{H}=\frac{h}{e^{2}\nu}$. The dashed line represents the universal resistivity asymptote $\rho^{univ}$. Inset: enlarged plot of the low-B $\rho$-data}
\end{center}
\end{figure}

\begin{figure}
\begin{center}
\includegraphics[scale=0.75]{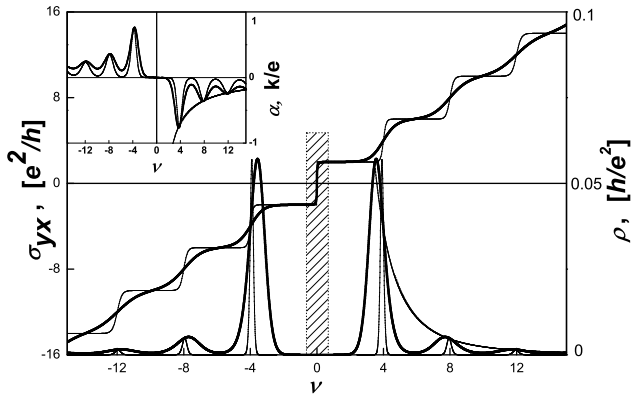} \caption[]{\label{Fig5} Hall conductivity $\sigma_{yx}$ and the total resistivity $\rho$ given
by Eq.(\ref{SLG_SIGMAyx}) and Eqs.(\ref{RESISTIVITY_HIGH_B},\ref{SLG_SIGMAyx},\ref{ALPHA_N_P}),
respectively, plotted as functions of the filling factor $\nu$ at different temperatures
$\vartheta_{1}=0.02,0.1$ for energy spectrum specified by
Eq.(\ref{SLG_SPECTRUM_B}). The dashed line corresponds to that in Fig.\ref{Fig4}. In the vicinity of the Dirac point (shaded area), the total resistivity is plotted in Fig.\ref{Fig6} for the split-off zeroth-LL energy spectrum. Inset: SLG thermoelectric power given by Eq.(\ref{ALPHA_N_P}) vs. the filling factor $\nu$ at $\vartheta_{1}=0.06,0.1$. The dashed line represents the asymptote $\alpha_{e} =-\frac{k}{e}\frac{4\ln2}{\nu}$.}
\end{center}
\end{figure}

Using Eq.(\ref{RESISTIVITY_HIGH_B},\ref{SLG_N},\ref{ALPHA_DEFINITION_ELECTRON}), in Fig.\ref{Fig4} we plot the
dependencies of the Hall resistivity $\rho_{H}=\frac{B}{Nec}$ and the total resistivity $\rho$ as a functions of the magnetic field. At low magnetic fields, $\rho_{H}=\frac{h}{e^{2}\nu} \sim B$. Then, in a quantizing magnetic field, the Hall resistivity exhibits the QH plateau at $\rho_{H}= \frac{h}{e^{2}}\frac{1}{4\mathcal{N}+2}$, accompanied by a plateau-to-plateau transition at the so-called critical filings, $\nu_{\mathcal{N}}=4\mathcal{N}$, when the Fermi energy coincides with the $\mathcal{N}$-th Landau level. Simultaneously, in the vicinity of the critical fillings $\nu_{\mathcal{N}}$, the total resistivity $\rho$ exhibits peaks. It is noteworthy that, at critical filings $\nu_{\mathcal{N}}$, the thermoelectric power, carrier density and, hence, resistivities given by Eq.(\ref{RESISTIVITY_HIGH_B}) approach the universal values $\alpha_{e}^{(\mathcal{N})}=-\frac{k}{e}\frac{4\ln 2}{\nu _{\mathcal{N}}}$, $\rho_{H}^{(\mathcal{N})}=\frac{h}{e^{2}}\nu _{\mathcal{N}}^{-1}$, and
$\rho^{(\mathcal{N})}=\rho_{H}^{(\mathcal{N})}(\alpha _e^{(\mathcal{N})})^{2}/L$, irrespective of
temperature. The dotted line in Fig.\ref{Fig4} represents the universal resistivity asymptote
$\rho^{univ}(\nu) =\frac{h}{e^{2}}\frac{3(4\ln2)^2}{\pi^{2}\nu^{3}}$ whose values at $\nu=\nu_{\mathcal{N}}$ correspond to
the universal resistivity values $\rho ^{(\mathcal{N})}$. Then, using  Eqs.(\ref{RESISTIVITY_HIGH_B},\ref{ALPHA_DEFINITION_ELECTRON},\ref{SLG_NS_LIFSHITZ}), we plot in the inset of Fig.\ref{Fig4} the low-B magnetoresistivity which exhibits onset of SdH oscillations
at $\nu^{-1} \sim \xi $. We emphasize that the SLG energy spectrum specified by Eq.(\ref{SLG_SPECTRUM_B})
suggests that zeroth-LL always remains occupied at $\mu > 0$, and, therefore, the Hall resistivity is
finite even in ultrahigh magnetic fields.

\subsection{ \label{sec: IQHE in SLG at fixed magnetic field}IQHE in SLG at fixed magnetic field}
We now examine the magneto-transport in SLG at a fixed magnetic field for an arbitrary $\mu$, i.e., in presence of electrons and holes. At a moment, we neglect the longitudinal conductivity of the 2D electrons i.e. $\sigma_{xx}=0$.  Using Eqs.(\ref{NP_DEFINITION},\ref{SLG_OMEGA_B}) and, in addition, combining the
electron(hole) carrier contributions, we finally obtain the Hall conductivity of monolayer graphene
\begin{equation}
\sigma_{yx}=\frac{4e^{2}}{h}\left[\sum \limits_{\mathcal{N}=1}^{\infty}\left[\mathcal{F}
\left(\frac{\varepsilon_{\mathcal{N}}-1}{\xi}\right)-
\mathcal{F}\left(\frac{\varepsilon_{\mathcal{N}}+1}{\xi}\right)\right]+\frac{\text{th} \left(\frac{1}{2\xi}
\right)}{2}\right ].
\label{SLG_SIGMAyx}
\end{equation}
According to Eq.(\ref{SIGMAxx},\ref{SLG_SIGMAyx}), the Hall conductivity is an odd function of
the magnetic field and Fermi energy. In Fig.\ref{Fig5}, the Hall conductivity is represented as a
function of the filling factor $\nu$. It is noteworthy that the dimensionless temperature embedded into Eq.(\ref{SLG_SIGMAyx}) depends on the filling factor. Indeed,  $\xi=\frac{kT}{\mu}=\frac{\vartheta_{1}}{\text{sign}(\nu)\sqrt{|\nu}|}$, where we introduced the dimensionless temperature at a fixed magnetic field, $\vartheta_{1}=\frac{\sqrt{2}kT\lambda_{B}}{\hbar \upsilon}$. For the typical $B=9$T, $\upsilon=8 \times 10^{7}$cm/s, and $T=4.2$K, we estimate $\vartheta_{1}=0.01$. Figure \ref{Fig5} demonstrates the sequence of the QH plateaus $\sigma_{yx}= \pm \frac{4e^{2}}{h}(\mathcal{N}+1/2)$ accompanied by the plateau-to-plateau transitions at the critical values $\sigma_{yx}^{(\mathcal{N})}=\frac{e^{2}}{h}\nu_{\mathcal{N}} $ of the Hall conductivity. In the vicinity of a certain plateau-to-plateau transition, $\Delta \nu=\nu-\nu_{\mathcal{N}}, \Delta \nu \ll \nu_{\mathcal{N}} $, the Hall conductivity
$\sigma_{yx}=\sigma_{yx}^{(\mathcal{N})}(1+\frac{\Delta \nu}{2\nu_{\mathcal{N}}^{3/2}\vartheta_{1}})$
exhibits the linear T-scaling. It is noteworthy that, in contrast to conventional 2D systems, the width of the
$\mathcal{N}$-th plateau-to-plateau transition in SLG increases as $ \nu_{\mathcal{N}}^{3/2} \sim N_{0}^{3/2}$.

Of particular interest is the behavior of the Hall conductivity in the vicinity of Dirac point. According to energy
spectrum given by Eq.(\ref{SLG_SPECTRUM_B}), the electron( hole ) zeroth-LLs are half-filled at $\mu=0$. Thus,
the Hall conductivity specified by Eq.(\ref{SLG_SIGMAyx}) exhibits a $4e^{2}/h$ step at DP  within the
ultra-narrow range of fillings $\nu \sim \vartheta_{1}^{2}$.

In order to complete our study of the magneto-transport in SLG for mixed state containing
both electrons and holes, we find the thermoelectric power
\begin{eqnarray}
\alpha=\frac{\alpha_{e}N+\alpha_{h}P}{N+P},
\label{ALPHA_N_P}\\
\alpha_{h}=-\alpha_{e}(-\mu)
\nonumber
\end{eqnarray}
where $\alpha_{h}$ is the thermoelectric power of holes. The thermoelectric power of monolayer graphene, specified by Eq.(\ref{ALPHA_N_P}), coincides with that for electrons and holes at $N \gg P$ and $N \ll P$, respectively. For strong magnetic fields, we use Eqs.(\ref{RESISTIVITY_HIGH_B},\ref{SLG_SIGMAyx},\ref{ALPHA_N_P}) and, plot in Fig.\ref{Fig5} the total resistivity against the filling factor. In agreement with experiment\cite{Novoselov04,Novoselov05}, the dependence $\rho(\nu)$ exhibits peaks in the vicinity of the critical filings $\nu_{\mathcal{N}}$, except for zeroth-LL-related peak at $\nu=0$. Below, we attempt to explain this discrepancy.

We argue that the magnetoresistivity vanishes near the Dirac point (see Fig.\ref{Fig5}) which is caused by the 4-fold degeneracy of the zeroth-LL filled by both electrons and holes. Indeed, only the zeroth-LL can make the main contribution to the electron $\Omega_{e}=2\Gamma \ln(1+e^{\frac{1}{\xi}})$
and hole $\Omega_{h}=\Omega_{e}(-\xi)$ thermodynamic potentials. After some algebra based
on Eqs.(\ref{ALPHA_N_P},\ref{NP_DEFINITION}), we can surprisingly
find that $\alpha \equiv 0$ (see the inset of Fig.\ref{Fig5}) within a wide range of Fermi energies when $\mid \nu \mid \leq 4$. Accordingly, the total resistivity vanishes within the above
$\nu$-range as well (see the shaded area in Fig.\ref{Fig5}).

Our results seem to contradict to the experimental observations\cite{Novoselov04,Novoselov05}, according to which a well-pronounced $\rho$-peak is observed at the Dirac point. We now demonstrate that the above discrepancy may be due to lifting of the four-fold degeneracy of zeroth-LL in real systems.

\subsection{ \label{sec: IQHE in SLG in the vicinity of the Dirac point}IQHE in SLG in the vicinity of the Dirac point}
In this section we resolve the magnetotransport problem in the vicinity of the Dirac point taking into account the zero-LL degeneracy lifting, and moreover assuming finite SLG conductivity $\sigma_{xx} \neq 0$.
The nature of the energy states in graphene at $\mathcal{N}=0$ LL has recently been under intense theoretical
and experimental \cite{Novoselov05,Cho08,Abanin07,Checkelsky08,Checkelsky09,Giesbers09,LZhang10,ZJiang07,YZhang06}
investigation, but still remains unresolved. Conventionally, the four-fold degeneracy of the zeroth-LL is associated with a two-fold degeneracy resulting from the spin symmetry and a two-fold degeneracy caused by sub-lattice
symmetry. We now make an attempt to distinguish the hierarchy of the split-off zeroth-LL on the basis of the available experimental data( see Table \ref{table}).

\begin{figure}
\begin{center}
\includegraphics[scale=0.75]{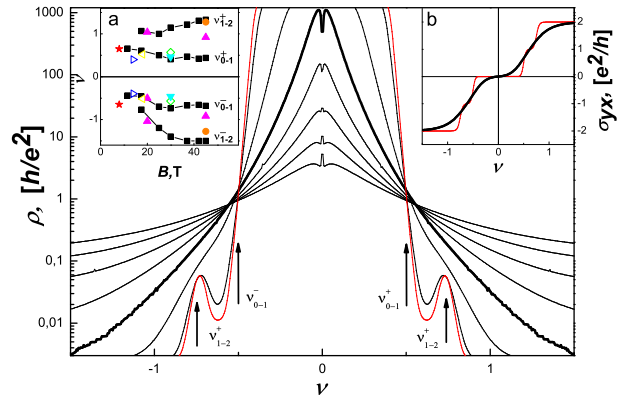} \caption[]{\label{Fig6} Evolution of the total resistivity
$\rho$ in the presence of zeroth-LL splitting at critical fillings $\nu^{\pm}_{0-1}=\pm 0.5$ and $\nu^{\pm}_{1-2}=\pm 0.75$
( corresponds to B=14T in inset a ) for dimensionless temperatures
$\vartheta_{1}=0.3,0.25,0.2,0.15,0.1,0.05,0.02,0.015$. SLG  conductivity $\sigma_{xx}=0.01 e^{2}/h$.
Inset a: Hierarchy of the split-off subbranches of
zeroth-LL. Direct observation of critical filling, based on T-independent points: $\triangleright$-\cite{Checkelsky08},
$\triangleleft$-\cite{LZhang10}, $\lozenge$-\cite{Giesbers09}. Indirect analysis of the critical filling
on the basis of $\hat{\rho}^{-1}$ data: $\bigstar$-\cite{Cho08}, $\blacktriangledown$-\cite{Abanin07},
$\blacksquare$-\cite{YZhang06}. Proximate analysis based on the position of the split-off $\rho$-peaks
$\blacktriangle$-\cite{ZJiang07}, $\bullet$-\cite{YZhang06}. Inset b: Hall conductivity at $\vartheta_{1}=0.1, 0.015$.}
\end{center}
\end{figure}

For moderately disordered samples at a fixed temperature and a fixed magnetic field $\leq 30$T, the density dependence of the magnetoresistivity exhibits a symmetric peak centered at DP. The height of the resistivity peak
increases as temperature is lowered. Then, the set of the density-dependent
resistivity isotherm curves (typically $0.5-50$K) clearly demonstrates two T-independent points $\rho \sim 0.5 h/e^{2}$ on the electron and hole sides of the resistivity peak. These points
lie at certain critical fillings $\sim \pm 0.5$( see open symbols in Fig.\ref{Fig6},inset "a"), which we
attribute to crossing of the electron($+$) and hole($-$) subbranches of the split-off zeroth-LL by the Fermi level. It is noteworthy that, usually, experimental $\hat{\rho}$-data are converted to give the inverse resistivity tensor $\hat{\rho}^{-1}$. At the above-mentioned critical fillings of $\sim \pm 0.5$, the longitudinal component of the inverse resistivity tensor $(\hat{\rho}^{-1})_{xx}$ also exhibits T-independent points, whereas the transverse component $(\hat{\rho}^{-1})_{xy}$ experiences a smooth
$(0-2)\frac{e^{2}}{\hbar}$ transition\cite{Cho08,Abanin07,Checkelsky08,Giesbers09,LZhang10}.
Actually, the $\hat{\rho}^{-1}$-data provide an alternative method for finding the critical fillings.
\begin{table}
\caption{\label{table}The device parameters used to evaluate the zeroth-LL splitting and the
magnetoresistivity at the critical fillings $\nu^{\pm}_{0-1}$ and $\nu^{\pm}_{1-2}$ in Fig.\ref{Fig6},inset "a"}
\begin{tabular}{@{}*{7}{l}}
\hline\hline $B$,T&$T$,K&$\mu$,1/T&$\rho^{(0-1)}/\rho^{(1-2)},\frac{h}{e^{2}}$&Ref.\\
\hline
   8&       2.3&   2& 0.6  &\cite{Cho08} \\
  14&   0.3-100& 1.3& 0.98 &\cite{Checkelsky08}\\
  18&    0.5-50& 0.6& 0.5  &\cite{LZhang10} \\
  20&       4.2&   2& 0.27/0.31&\cite{ZJiang07} \\
  30&         4&    & 0.42 &\cite{Abanin07} \\
  - &   0.4-150&   1& 0.38 &\cite{Giesbers09} \\
  45&   4.2-294&  2& -/0.3&\cite{ZJiang07} \\
  - &       1.4&   5& -/0.2&\cite{YZhang06} \\
9-45&       0.03&  5&      &\cite{YZhang06} \\
 \hline\hline
\end{tabular}
\end{table}

Further progress related to the zeroth-LL splitting problem has been made  for high-mobility
samples at low temperatures\cite{YZhang06}. Surprisingly, at ultra-low temperatures ($\sim 30$mK) the previously
reported\cite{Checkelsky08,Giesbers09} smooth transition $(0-2)\frac{e^{2}}{\hbar}$ in $(\hat{\rho}^{-1})_{xy}$
turns into a sequence( $(0-1)\frac{e^{2}}{\hbar}$ and $(1-2)\frac{e^{2}}{\hbar}$) of two sharp transitions.
These transitions occur at certain critical fillings $\nu^{\pm}_{0-1}$ and $\nu^{\pm}_{1-2}$
associated with the crossing of the electron($+$) and hole($-$) related lowest(0-1) and highest(1-2)
zeroth-LL split subbranches by Fermi level. In Fig.\ref{Fig6},inset "a" we plot a fun diagram of split-off zeroth-LL critical fillings and, moreover, add the data of \cite{ZJiang07,YZhang06} for ultra-high fields $\leq 45$T at which the resistivity exhibits well defined split-off zeroth-LL peaks near the DP. For the latter high-field case, the critical fillings were determined by careful analysis of the split-off resistivity peak positions. However, exact definition of the zeroth-LL subbranches( spin and(or) valley ) remains unclear. The available tilt-field experimental data\cite{ZJiang07,YZhang06} point to only the spin-independent origin of zeroth-LL splitting.

Of particular interest is the behavior of the density-dependent resistivity $\rho$ measured in the vicinity of the Dirac point at a fixed temperature for different applied magnetic fields. When the magnetic field grows ($B\leq 31$T), both the height and width of the DP resistivity peak increase\cite{Checkelsky09}. Moreover, at the Dirac point $\rho(B)$ dependence exhibits\cite{Checkelsky08,Checkelsky09} the abrupt rise at a certain critical magnetic field. The higher the sample mobility at $B=0$( i.e. the smaller the offset gate bias ), the
lower the critical magnetic field. Intriguingly, the expected divergence of the DP magnetoresistivity is not observed\cite{JPoumirol10} for high-disorder samples $ \sim 10^{3}\frac{cm^2}{Vs}$ up to $B=57$T. We
argue that these effects are highly expected within our scenario. It will be recalled that the resistivity is a
universal function of the filling factor $\nu$ and the dimensionless temperature $\vartheta_{1} \sim kT/\sqrt{B}$.
Consequently, the observed\cite{Checkelsky08}  enhancement of the resistivity, caused by the increase in the magnetic field at $T$=const, may be similar to that reported \cite{Checkelsky09,LZhang10,Giesbers09} for the $T \rightarrow 0$ sweep at $B$=const.

To compare our results with the experiment\cite{Checkelsky08,Checkelsky09}, we plot in Fig.\ref{Fig6} the
transport coefficients $\rho, \sigma_{yx}$ in the vicinity of the Dirac point on the assumption that the critical fillings have values $\nu^{\pm}_{0-1}=\pm 0.5$ and $\nu^{\pm}_{1-2}=\pm 0.75$ deduced from Fig.\ref{Fig6},inset "a" for B=14T. Then, we set the constant longitudinal conductivity $\sigma_{xx}=0.01e^{2}/h$, which provides a reasonable value for the sample resistivity $\sigma_{xx}^{-1} \sim 2.5$M$\Omega$ at the Dirac point. It is noteworthy that, at $T\leq 5$K the $\rho(\nu)$-curves found to be indistinguishable in \cite{LZhang10}, thus we can roughly
estimate the typical LL width as $\sim 5$K. Finally, we set in Fig.\ref{Fig6} $\vartheta_{1}=0.015-0.3$ for
the temperature range $T=10-225$K. At high temperatures, the T-independent point at the electron(hole) sides is clearly seen. By contrast, at $T \rightarrow 0$, the  electron(hole) zeroth-LL subbranches become resolved. The ultranarrow peak(deep) $\sigma_{xx}^{-1}$ at the Dirac point is attributed to 2D graphene resistivity when $\sigma_{yx} \ll \sigma_{xx}$. We emphasize that dependencies of the $\rho(B), \rho(T)$ at the Dirac point $\nu=0$ can provide the unique knowledge regarding the own magnetoresistivity $\rho_{xx}$ of 2D graphene flake.

It is instructive to compare the resistivities at the critical fillings $\nu_{(0-1)},\nu_{(1-2)}$  with
those reported in experiment( see Table \ref{table} ). For half-filled lower(0-1) and higher(1-2) zeroth-LL subbranch we obtain the electron densities $1/2\Gamma$ and $3/2\Gamma$, respectively. Then, the entropy $S_{e}=k\Gamma \ln2$
is the same in both cases. The critical resistivities are $\rho^{(0-1)}=\frac{3(\ln2)^2h}{\pi^{2}0.5^{3}e^{2}}=1.17
\frac {h}{e^{2}}$  and $\rho^{(1-2)}=0.043\frac {h}{e^{2}}$, respectively. According to Fig.\ref{Fig6}, only the critical point at $\nu^{\pm}_{0-1}$ persists at
elevated temperatures, thus, the respective critical
resistivity $\rho^{(0-1)}$ is consistent with that ($\sim 0.5\frac{h}{e^{2}}$) observed in
experiment (see Table \ref{table}).

\begin{figure}
\begin{center}
\includegraphics[scale=0.75]{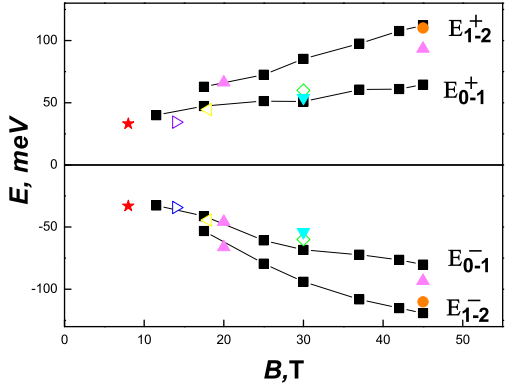} \caption[]{\label{Fig7} The energy of split-off zeroth-LL subbranches vs magnetic field in SLG. The data point notation is the same that in Fig.\ref{Fig6}, inset "a".}
\end{center}
\end{figure}

Finally, we calculate the absolute value of energy $E=\hbar \upsilon \sqrt{\pi \Gamma \nu}$ or in dimension units $E(\text{meV})=14,5\sqrt{\nu B(\text{T})}$ related to zero LL split sublevels. In Fig.\ref{Fig7} we re-plot the results represented in Fig.\ref{Fig6}, inset "a" in terms of zero LL split energy vs magnetic field.

\subsection{\label{sec: IQHE in bilayer graphene} IQHE in bilayer graphene}
We now analyze the case of bilayer graphene(BLG) magnetotransport, which differs insignificantly from that
considered above for monolayer graphene. With our previous notations, the BLG energy
spectrum\cite{Falko06} is given by
\begin{equation}
E_{\mathcal{N}}^{\pm}=\pm E_{\mathcal{N}}=\pm \hbar \omega_{c}\sqrt{\mathcal{N}(\mathcal{N}-1)},
\label{BLG_SPECTRUM_B}
\end{equation}
where $\omega_{c}=\frac{eB}{mc}$ is the cyclotron frequency; $m=\frac{\gamma}{2\upsilon^{2}}=0.054m_{e}$,
effective mass; $\gamma=0.39$eV, interlayer coupling constant. For the general
case of an arbitrary interlayer coupling, the BLG spectrum is described\cite{Vasilopoulos08} by Eq.(\ref{BLG-SLG_SPECTRUM_B}) (see Appendix \ref{sec:Bilayer graphene spectrum}).
\begin{figure}
\begin{center}
\includegraphics[scale=0.75]{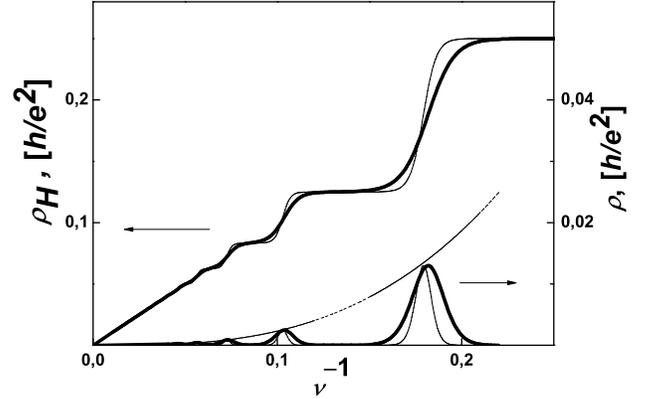} \caption[]{\label{Fig8} QHE in BLG at fixed $\mu>0$ for $\xi=0.2,0.4$. The dashed lines correspond to universal resistivity asymptote $\rho^{univ}$.}
\end{center}
\end{figure}

The electron- and hole-related parts of the BLG thermodynamic potential are
\begin{eqnarray}
\Omega_{e}=-kT \Gamma \left[4\sum \limits_{N=2}^{\infty}\ln
(1+e^{\frac{\mu -E_{\mathcal{N}}}{kT}})+ 4\ln(1+e^{\frac{\mu}{kT}})\right],
\label{BLG_OMEGA_ELECTRON}\\
\Omega_{h}=\Omega_{e}(-\mu, T),
\nonumber
\end{eqnarray}
where we take into account the four-fold degeneracy( spin+valley ) for $\mathcal{N} \geq 2$-index LLs. Then,
based on the energy spectrum given by Eq.(\ref{BLG_SPECTRUM_B}) we take into account the total
8-fold degeneracy of the zero-energy LL due to the spin, valley, and $\mathcal{N}=0,1$ orbital LL degeneracies.

Similarly to the SLG case, we introduce the dimensionless electron energy spectrum as
$\varepsilon_{\mathcal{N}}= \frac{4}{\nu}\sqrt{\mathcal{N}(\mathcal{N}-1)}$. Using Eqs.(\ref{BLG_OMEGA_ELECTRON}, \ref{NP_DEFINITION},\ref{ALPHA_DEFINITION_ELECTRON},\ref{RESISTIVITY_HIGH_B}) we plot in Fig.\ref{Fig8} the magnetic field dependence of the Hall resistivity $\rho_{H}=\frac{B}{Nec}$ and
of the total resistivity $\rho$. The dashed line represents the resistivity asymptote $\rho^{univ}(\nu)$ whose values at $\nu_{\mathcal{N}}=4\sqrt{\mathcal{N}(\mathcal{N}-1)},\mathcal{N}\geq 2$ gives the respective universal values when the Fermi energy coincides with the $\mathcal{N}$-th Landau level.
\begin{figure}
\begin{center}
\includegraphics[scale=0.75]{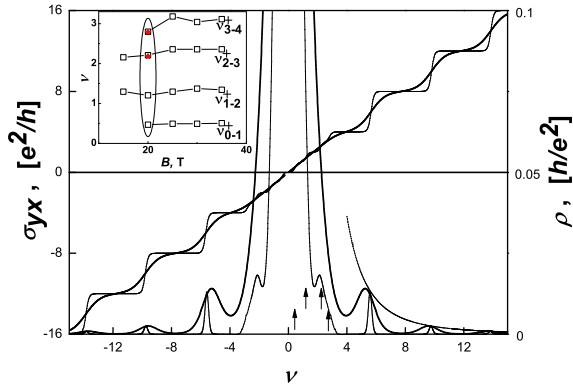} \caption[]{\label{Fig9} QHE in BLG at fixed magnetic field $B=20$T, dimensionless temperature $\vartheta_{2}=0.5,0.2$ and $\sigma_{xx}=0.01 e^{2}/h$. The dashed line
correspond to universal resistivity asymptote $\rho^{univ}$. Inset: magnetic filed dependence of the electron(+) split-off $\mathcal{N}=0,1$ LL subbranch fillings. The arrows in the main panel corresponds to fillings at $B=20$T in inset.}
\end{center}
\end{figure}

We now extend our previous results for the bipolar magneto-transport in SLG at $B$=const for the present case of bilayer graphene.
Using Eqs.(\ref{BLG_OMEGA_ELECTRON},\ref{NP_DEFINITION},\ref{SIGMAxx}), we obtain for the BLG Hall conductivity
\begin{equation}
\sigma_{yx}=\frac{4e^{2}}{h}\left[\sum \limits_{\mathcal{N}=1}^{\infty}\left[\mathcal{F}
\left(\frac{\varepsilon_{\mathcal{N}}-1}{\xi}\right)-
\mathcal{F}\left(\frac{\varepsilon_{\mathcal{N}}+1}{\xi}\right)\right]+\frac{\text{th} \left(\frac{1}{2\xi}
\right)}{4}\right ].
\label{BLG_SIGMAyx}
\end{equation}
It is worthwhile to mention that the dimensionless temperature embedded into
Eq.(\ref{BLG_SIGMAyx}) can be re-written as $\xi=\frac{\vartheta_{2}}{\nu}$, where we introduced the
dimensionless temperature at a fixed magnetic field, $\vartheta_{2}=\frac{4kT}{\hbar \omega_{c}}$. In Fig.\ref{Fig9}, the Hall conductivity is represented as a function of the filling factor $\nu$. Assuming that the charge transport is dissipationless in strong magnetic fields, we
employ Eqs.(\ref{RESISTIVITY_HIGH_B},\ref{NP_DEFINITION},\ref{ALPHA_N_P},\ref{BLG_SIGMAyx})
and plot the total resistivity in Fig.\ref{Fig9}. It exhibits peaks in the vicinity of the critical
filings $\nu_{\mathcal{N}}$ in agreement with experimental findings\cite{Novoselov06}.

In order to analyze the $\rho$-behavior in the vicinity of the Dirac point, we use the available
experimental data\cite{Zhao10} and identify the split-off subbranches of the zero-energy LLs($\mathcal{N}=0,1$). Our result is represented in Fig.\ref{Fig9},inset for electron branch(marked as "+") of the BLG spectrum. In strong magnetic fields the electron-related zero-energy LLs splits into four sub-levels $\nu_{0-1}^{+},\nu_{1-2}^{+},\nu_{2-3}^{+},\nu_{3-4}^{+}$ shown in Fig.\ref{Fig9},inset. Similar to monolayer graphene case the resistivity peak at the Dirac point is caused by split-off subbranches of the zero-energy LLs.

\begin{figure}
\begin{center}
\includegraphics[scale=0.75]{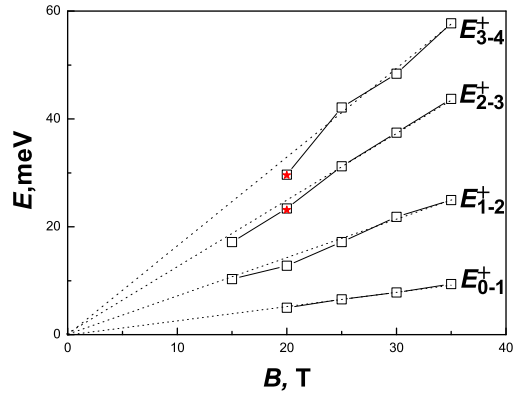} \caption[]{\label{Fig10} BLG hierarchy of the
split-off electron subbranches of $\mathcal{N}=0,1$ LL. The critical fillings are extracted from $\hat{\rho}^{-1}$-data
$\Box$-\cite{Zhao10}; analyzing the position of the split-off $\rho$-peaks $\bigstar$-\cite{Zhao10}.}
\end{center}
\end{figure}

Finally, we calculate the absolute value of energy $E=\frac{\pi \hbar^{2}\upsilon^{2}}{\gamma}\Gamma \nu$ or in dimensional units $E(\text{meV})=0.53\nu B(\text{T})$ related to zero-energy split LL sublevels. For electrons the result is shown in Fig.\ref{Fig10}. As expected, the $\mathcal{N}=0,1$ LL splitting disappears at zero magnetic field.

\section{\label{sec:conclusion}Conclusions}
In conclusion, we studied the Quantum Hall effect in mono- and bi-layer graphene. The transverse and longitudinal resistivities were found as a universal function of the filling factor and temperature. Based on experimental data, we found the B-dependence of zeroth-LL split subbranches and, in addition, resolved the magneto-transport problem in the vicinity of the Dirac point.
\section{\label{sec:Appendix} Appendix}
\subsection{\label{sec:Lifshitz-Kosevich formalism} Lifshitz-Kosevich formalism}
\label{Lifshitz-Kosevich formalism}
Using the conventional Poisson formulae
\begin{equation}
\sum\limits_{m_{0}}^{\infty }\varphi (\mathcal{N})=\int\limits_{a}^{\infty
}\varphi (\mathcal{N})d\mathcal{N}+2\text{Re}\sum\limits_{k=1}^{\infty
}\int\limits_{a}^{\infty }\varphi (N)e^{2\pi ik\mathcal{N}}d\mathcal{N},
\label{POISSON}
\end{equation}
where $m_{0}-1<a<m_{0}$ ($m_{0}$ is lower limit of summation), the thermodynamic potential can be represented as the sum $ \Omega_{e}=\Omega_{-}+\Omega_{\sim}$ of the zero-field and oscillating parts as
\begin{eqnarray}
\Omega_{-}=-N_{0} \mu \xi^{3} F_{2}(1/\xi),
\label{SLG_OMEGA1_LIFSHITZ} \\
\Omega_{\sim}=-8\frac{ N_{0} \mu \xi}{\nu} \times {\text Re}
\sum\limits_{k=1}^{\infty }\int\limits_{0}^{\infty }e^{2\pi ik\mathcal{N}}
\ln \left( 1+e^{\frac{1-\varepsilon_{\mathcal{N}}}{\xi}} \right )d \mathcal{N},
\nonumber
\end{eqnarray}
After a simple integration by parts, the oscillating term has the form
\begin{equation}
\Omega_{\sim}=8\frac{ N_{0} \mu }{\nu}  \times \text{Re} \sum\limits_{k=1}^{\infty
}\frac{1}{2\pi i k } \int \limits_{0}^{\infty } \frac{e^{2\pi
ik\mathcal{N}(\varepsilon)}}{1+e^{\frac{\varepsilon-1}{\xi}}}d\varepsilon
\label{SLG_OMEGA2_LIFSHITZ}
\end{equation}

It should be noted that the integrand in
Eq.(\ref{SLG_OMEGA2_LIFSHITZ}) is a rapidly oscillating function, which is,
in addition, strongly damped at $\varepsilon
> 1$. A major part of the integral results from the
energy range close to the Fermi energy, when $\varepsilon \sim 1$.
Therefore, $\mathcal{N}(\varepsilon)=\frac{\varepsilon^{2}\nu}{4}$ can be regarded as smooth
functions of energy, and, hence, can be re-written as
\begin{equation}
\mathcal{N} \mid_{\varepsilon=1}=\frac{\nu}{4}+\frac{\nu}{2}(\varepsilon-1).
\label{SLG_N(E)}
\end{equation}
Under the above assumption, we can
change the lower limit of integration to $-\infty$ and then use
the textbook expression $\int\limits_{-\infty}^{\infty}\frac{e^{i
k y}}{1+e^{y}}dy=\frac{-i\pi}{\sinh(\pi k)}$ for the integral of
the above type. Finally, we have for the thermodynamic potential
\begin{equation}
\Omega_{e}=\Omega_{-} + 4 \frac{ N_{0} \mu \xi}{\nu}
\sum\limits_{k=1}^{\infty }\frac{\cos(\pi k\nu/2)}{k \sinh{r_{k}}}
\label{SLG_OMEGA3_LIFSHITZ}
\end{equation}
where $r_{k}=\pi^{2} k\nu \xi$ is a dimensionless parameter related to T-damping
of the oscillations $\sim \cos(\pi k\nu/2)$. Using the conventional
thermodynamic definition and Eq.(\ref{SLG_OMEGA2_LIFSHITZ}), we find both the entropy
and the density of Dirac electrons, valid at low temperatures and weak
magnetic fields $\xi, \nu^{-1} \ll 1$:
\begin{eqnarray}
N=N_{T}+4N_{0}\pi \xi  \sum
\limits_{k=1}^{\infty} \frac{\sin \left(\frac{\pi k\nu}{2}\right )+\pi \xi \cos\left(\frac{\pi k\nu}{2}\right )
\coth(r_{k})}{\sinh(r_{k})},
\label{SLG_NS_LIFSHITZ} \\
S_{e}=S-4k_{B}N_{0} \sum \limits_{k=1}^{\infty }\frac{1}{k\nu} \cos\left(\frac{\pi k\nu}{2}\right )\Phi
(r_{k}), \nonumber
\end{eqnarray}
where $N_{T}=N_{0}2\xi^{2} F_{1}(1/\xi )$ and
$S_{T}= 2k_{B}N_{0}(\frac{3}{2}\xi^{2} F_{2}(1/\xi)-\xi F_{1}(1/\xi))$ are the
carrier density and entropy at $B=0$, respectively; and, $\Phi
(z)=\frac{1-z\coth (z)}{\sinh (z)}$ is the form-factor.

\subsection{\label{sec:Bilayer graphene spectrum}Spectrum of bilayer graphene at finite interlayer coupling}
\label{BLG-SLG}

At B=0, the spectrum of bilayer graphene spectrum was found in Ref.\cite{Falko06} as
\begin{equation}
E(k)=\pm \frac{1}{2}\gamma\left[\sqrt{1+4(\upsilon \hbar k)^{2}/\gamma^{2}} -1 \right].
\label{BLG-SLG_SPECTRUM_B0}
\end{equation}
For clarity, we further consider the electron-related part("+") of the spectrum. For a strong inter-layer
coupling $\frac{2\upsilon \hbar k}{\gamma} \ll 1$, the BLG spectrum $E(k)= \frac{\hbar^{2}k^{2}}{2m}$
coincides with that known for regular 2DEG systems. In the opposite case of a weak interlayer
coupling, i.e., a high carrier momentum $\frac{2\upsilon \hbar k}{\gamma} \gg 1$, Eq.(\ref{BLG-SLG_SPECTRUM_B0})
reproduces the linear Dirac spectrum $E(k)=\hbar \upsilon k$. For an arbitrary $\gamma$, we can find the BLG
density of states $D(E)=\frac{2E+\gamma}{\pi \hbar ^{2}\upsilon^{2}}$ and then calculate the electron
density for the actual $T \rightarrow 0$ case as
\begin{equation}
N=N_{0}\left(1+\frac{\gamma}{\mu}\right ).
\label{BLG-SLG_N_B0}
\end{equation}
For strong(weak) inter-layer coupling, Eq.(\ref{BLG-SLG_N_B0}) yields the relationship for the Fermi energy, $N=N_{0}\frac{\gamma}{\mu}=\frac{2m\mu}{\pi \hbar^{2}}$, known for the conventional 2DEG and that ( $N=N_{0}$) for Dirac electrons, respectively. The transition from the Fermi-like spectrum to that of the Dirac type occurs
at $\frac{2\upsilon \hbar k}{\gamma} \sim 1$, which yields the critical electron density
$N^{*}=\frac{\gamma^{2}}{4\pi \hbar^{2} \upsilon^{2}}$. It is noteworthy that Eq.(\ref{BLG-SLG_N_B0}) makes it possible to find explicitly the Fermi energy $\mu(N)=\gamma(\sqrt{1+N/N^{*}}-1)/2$. The BLG-to-SLG
transition of the energy spectrum occurs at $N=N^{*}$, which corresponds to the critical Fermi energy $\mu^{*}=\gamma(\sqrt{2}-1)/2$.
Assuming that $\gamma_{ñ}=0.39$eV and $\upsilon = 8 \times 10^{7}$ cm/s, we can obtain the following estimates
$N^{*}=4.36 \times 10^{12}$ cm$^{-2}$\cite{Falko06} and $\mu^{*}=0.08$eV.

\begin{figure}
\begin{center}
\includegraphics[scale=0.75]{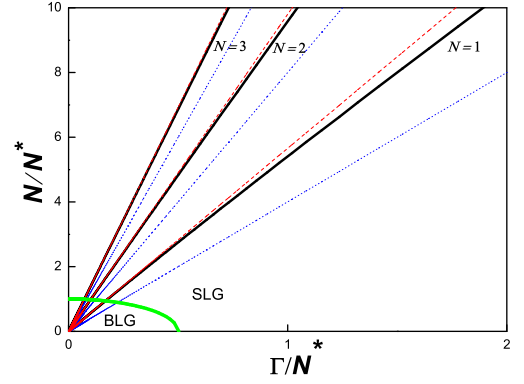} \caption[]{\label{Fig11} Phase diagram of the BLG electron
states for the lowest($\mathcal{N}=1,2,3$) LLs and an arbitrary inter-layer coupling. The red and blue lines are
related to the BLG($\nu=4\sqrt{\mathcal{N}(\mathcal{N}-1)}$) and SLG($\nu=4\mathcal{N}$) energy states,
respectively. The green line represents the BLG-to-SLG energy phase separator.}
\end{center}
\end{figure}

We emphasize that, in a strong magnetic field, the inter-layer coupling may play an important role as well. For arbitrary $\gamma$, the LL spectrum of the bilayer graphene was derived in Ref.\cite{Vasilopoulos08}
\begin{equation}
E_{\mathcal{N}}=\frac{\gamma}{\sqrt{2\beta}}\left[ \beta-1+2\mathcal{N}-\sqrt{(\beta-1)^{2}+4\mathcal{N}\beta}
\right]^{1/2}.
\label{BLG-SLG_SPECTRUM_B}
\end{equation}
Here, we introduce the inter-layer coupling parameter $\beta=N^{*}/\Gamma$. For the ultimate case of
SLG($\beta=0$) and BLG($\beta \rightarrow \infty$), it is possible to easily reconstruct
the discrete LL energy spectra specified by Eq.(\ref{SLG_SPECTRUM_B}) and Eq.(\ref{BLG_SPECTRUM_B}),
respectively. The BLG-to-SLG change in the energy spectrum occurs at $\beta \sim 2(2\mathcal{N}-1)$.
The higher the LL index, the lower the magnetic field at which the BLG-to-SLG transition occurs.

We can now find the phase diagram of BLG electron states $N(\Gamma)$ for an arbitrary inter-layer
coupling. Using the condition $E_{\mathcal{N}}=\mu(N)$ related to the Fermi level crossing the
$\mathcal{N}$-th LL and Eqs.(\ref{BLG-SLG_N_B0},\ref{BLG-SLG_SPECTRUM_B}), we calculate the phase
diagram( see Fig.\ref{Fig11}). Since both axes are scaled in density units $N^{*}$, the phase diagram
is valid for an arbitrary inter-layer coupling $\gamma$. With increasing magnetic field strength, the initial
BLG energy state( with a certain LL index ) is transformed to that belonging to SLG. We can easily find the
BLG-SLG phase separator $N/N^{*}=\sqrt{1-(2\Gamma/N^{*}})^{2}$ represented by the green line in Fig.\ref{Fig11}.
As expected, at $B \rightarrow 0$, the phase diagram demonstrates a BLG-to-SLG transition at $N=N^{*}$. It will be
recalled that, in a real BLG system, $N^{*}=4.36 \times 10^{12}$ cm$^{-2}$, and, therefore, the magnetic field
scale of the BLG-to-SLG transition (i.e. $\Gamma \sim N^{*}/2$) is on the order of $B\sim 90$T. We conclude that, for the actual experimental range $B=0-40$T the BLG spectrum can be well described by the simplified Eq.(\ref{BLG_SPECTRUM_B}).

\end{document}